\documentclass[twocolumn,showpacs,preprintnumbers,amsmath,amssymb]{revtex4}
\usepackage{graphicx}% Include figure files
\usepackage{dcolumn}% Align table columns on decimal point
\usepackage{bm}% bold math

\begin{document}
\preprint{APS/123-QED}
\title{Non linear excess conductivity of Bi$_2$Sr$_2$Ca$_{n-1}$Cu$_n$O$_{2n+4+x}$ ($n$ = 1, 2) thin films}
%\subtitle{} \titlerunning{running title}
\author{L. Fruchter}
\author{I. Sfar}
\altaffiliation[Also at ]{L.P.M.C., D\'{e}partement de Physique,
Facult\'{e} des Sciences de Tunis, campus universitaire 1060
Tunis, Tunisia.}
\author{F. Bouquet, Z.Z. Li}
\author {H. Raffy}
\affiliation{Laboratoire de Physique des Solides, C.N.R.S.
Universit\'{e} Paris-Sud, 91405 Orsay cedex, France}
\date{\today}
\begin{abstract}
The suppression of excess conductivity with electric field is
studied for Bi$_2$Sr$_2$Ca$_{n-1}$Cu$_n$O$_{2n+4+x}$ ($n$ = 1, 2)
thin films. A pulse-probe technique is used, which allows for  an
estimate of the sample temperature. The characteristic electric
field for fluctuations suppression is found well below the
expected value for all samples. For the $n=1$ material, a scaling
of the excess conductivity with electric field and temperature is
obtained, similar to the scaling under strong magnetic field.
\end{abstract}

\maketitle

\section{Introduction}

High-$T_{c}$ superconductors exhibit strong superconducting
fluctuations at the superconducting temperature, as a result of
the high ratio of the transition temperature to the condensation
energy - essentially due to their short coherence length. Such
fluctuations have been extensively studied in the weak and strong
fluctuation limits (respectively the Gaussian and the critical
regime). There is now a rising interest in the possibility to
suppress superconductivity by the application of an electric
field, and the study of the electric field dependence of the
fluctuations appears as a first step towards this goal. Early
theoretical work has focused on the reduction of the excess
conductivity in the Gaussian regime, by the application of a
transport current approaching the depairing current (non linear
conductivity). The electric field dependence of the
Aslamazov-Larkin term\cite{aslamazov68} was obtained in
Refs.~\cite{hurault69,schmidt69,tsuzuki69,gorkov70}, and the one
of the additional 'anomalous' Maki-Thompson term\cite{thompson70}
in Ref.~\cite{maki71}. Experimental results on Al films
demonstrated the suppression of fluctuations by the electric
field in the isotropic two-dimensional
limit\cite{kajimura70,kajimura71}. The theory, based on a
Langevin equation for the order parameter, was extended to
\textit{below} $T_{c}$ in Ref.~\cite{kajimura71b}. Later, several
works considered the contribution of the interaction of
fluctuations in the critical region and, recently, the expression
for the fluctuation conductivity in the transition region was
revisited and the similarity of the results with those obtained
for the broadening of the transition under a strong magnetic
field was pointed out (see Ref.~\cite{puica03} and refs therein).
Such experiments on high-$T_{c}$ superconductors are difficult,
as they require high current densities, which generally induce an
uncontrolled heating of the sample. As a result, only few
experimental works have investigated the effect of the electrical
field on the resistive transition of the cuprates. In
Ref.~\cite{soret93}, a field-induced crossover between a
three-dimensional and two-dimensional behavior was reported for a
YBa$_{2}$Cu$_{3}$O$_{7-x}$ single crystal, in the temperature
range $\epsilon = (T-T_{c})/T_{c} < 2\;10^{-2}$. In
Ref.~\cite{gorlova95}, the non-linear resistivity in the range
$0<\epsilon<7\;10^{-2}$ was used to obtained the characteristic
depairing field of Bi$_2$Sr$_2$CaCu$_2$O$_{8+x}$ single crystal.
It was found to be about 10 times smaller than the expected value
and its temperature dependence was clearly different from the
$\epsilon^{3/2}$ dependence predicted in Ref.~\cite{schmidt69}.

In this contribution, we compare the excess conductivity
suppression for this system with the one in
Bi$_2$Sr$_2$CuO$_{6+x}$, where the investigation of temperature
intervals as large as $-0.2 < \epsilon < 0.6$ was possible due to
an electrical critical field more than one order of magnitude
smaller.

\section{Experiment}

The measurements were carried out on
Bi$_2$Sr$_2$Ca$_{n-1}$Cu$_n$O$_{2n+4+x}$ ($n$ = 1, 2) epitaxial,
c-axis oriented thin films. They were grown on heated SrTiO$_3$
substrates, by reactive \textit{rf} sputtering with an oxygen
rich plasma (Ref.~\cite{li93} and refs therein). After deposition
of typically 2500~\AA{} thick films, Au contacts were sputtered,
and the samples were patterned in the four contact transport
geometry, with a current carrying strip of typical width and
length ($L$) of 80~$\mu$m and 200~$\mu$m.
%
% Pas totalement sur de ce changement, je n'avais pas vu
% que L etait utilise par la suite
%
The annealing of the samples at low temperature was used to set
the doping state, which was determined from the normal state
resistivity temperature dependence\cite{kon00}. The thin film
with $n=2$ (sample II) was overdoped. A thin film with $n=1$ was
studied for two different doping states: an overdoped one (sample
Ia) and an underdoped one (sample Ib). Two other samples with
$n=1$ (samples Ic and Id), respectively close to optimal doping
and strongly overdoped, were also investigated. The sample
parameters are summarized in Table~\ref{table}.

The resistivity in the limit of vanishing current density was
obtained using a 10~$\mu$A \textit{ac} current and a lock-in
amplifier detection. The investigation for large current
densities was carried out using the pulsed current technique. The
current pulses where 10~$\mu$s long, with a repetition rate
$\tau=10^{-4}$ and the maximum current value was 100~mA. Both the
voltage on a reference resistor fed with the measuring current
and the one on the sample, after amplification by home-made
amplifiers, were recorded, using a $2.5$~MHz, 16 bits digital
acquisition card. The measurement of the resistance, obtained
from the ratio of the two voltages, was linear within 0.3\% over
the current range. The non linearity was corrected from the raw
measurements. Evaluating the temperature increase due to the
measuring current is crucial for the high current density
transport experiments, as an apparent decrease of the excess
conductivity with current may occur, due to the sample
temperature increase only. A reliable determination of the non
linear excess conductivity requires a method for measuring the
sample temperature. For this purpose, we modified the technique
to a 'pulse-probe' one: after the completion of the main 10
$\mu$s pulse with current density $J$ and the measurement at the
end of this pulse, a probe pulse with current $J_s \ll J$ is
performed immediately after. A typical secondary current pulse
was $J_s \simeq J/50$. Neglecting the sample heating by current
$J_s$, the value of the resistance during the secondary pulse
provides a measure for the sample temperature about 1~$\mu$s
after the main pulse, using the ac resistivity data as a
thermometer. Such a method does not, however, catch the fast
temperature relaxation which occurs between the main and the
probe pulses. This relaxation, with a typical relaxation time in
the ns range\cite{carr90}, may be estimated as $\Delta T \simeq
R\,i^2/A\,\lambda$, where $A$ is the sample surface and
$\lambda\simeq10^7$ WK$^{-1}$m$^{-2}$ is the boundary
conductance, yielding $\Delta T \simeq 0.01$~K and 0.1~K for
samples with $n=1$ and $n=2$ respectively.

\section{results and discussion}
\label{results}

\begin{figure}
\resizebox{0.5\textwidth}{!}{%
\includegraphics{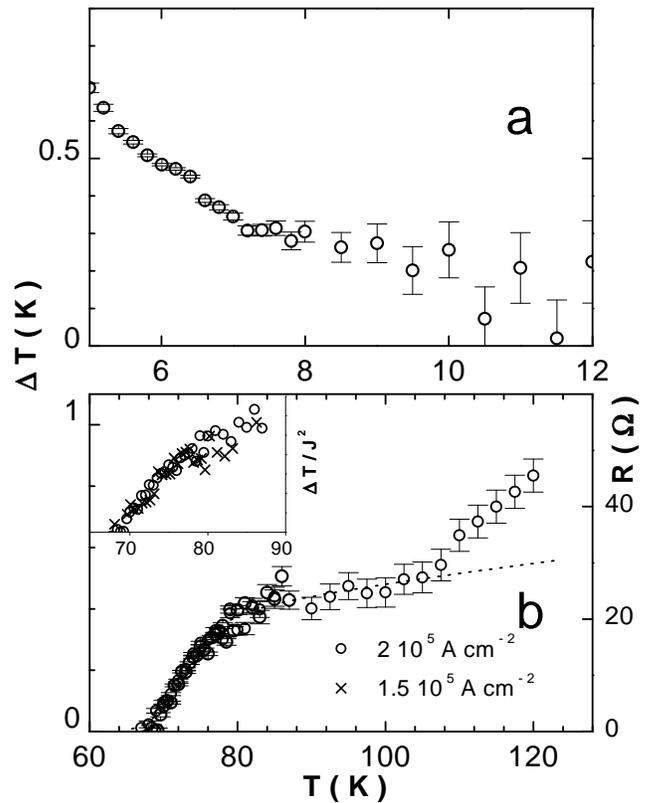}
} \caption{Sample Ia (a) and sample II (b): temperature difference
between the sample holder and the film, as measured by the probe
pulse.} \label{heating}
\end{figure}

\begin{figure}
\resizebox{0.5\textwidth}{!}{%
\includegraphics{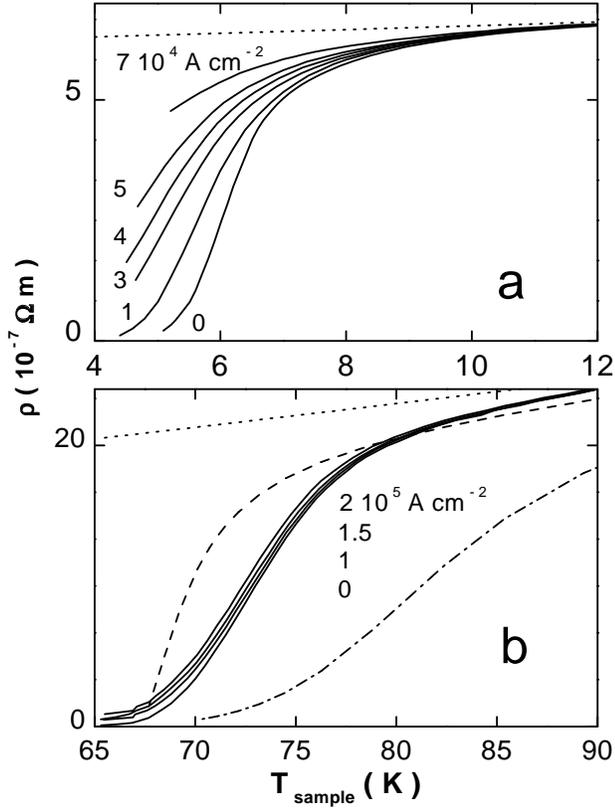}
} \caption{Sample Ia (a) and sample II (b): resistivity vs sample
temperature. Dotted line: the extrapolated normal state
resistivity $a + b T^\alpha$; dashed line: Hartree approximation
using the parameters in Ref.~\cite{livanov97}; dashed dotted line:
Hartree approximation using the raw superconducting parameters.}
\label{resistivity}
\end{figure}

The temperature increase for the over-doped
%  120303mmb
Bi$_2$Sr$_2$CaCu$_2$O$_{8+x}$ thin film (sample II) is shown in
Fig.~\ref{heating}b. The resistance of the current contacts for
this sample was $20\;\Omega$. As can be seen from the comparison
with the resistance of the sample and, also, from the scaling of
the sample heating with $J^2$, the data may be described using
the simple evaluation: $(T_{sample}-T_{experiment}) \propto
R(T_0,J)\;J^2$ below $\sim 100$ K. This demonstrates that the
strip temperature is essentially determined by the heating power
in the film constriction rather than by the one dissipated in the
current contacts. Using the probe value as the actual sample
temperature, a slight broadening of the transition remains which
must be attributed to the transport current
(Fig.~\ref{resistivity}b).

\begin{figure}
\resizebox{0.5\textwidth}{!}{%
\includegraphics{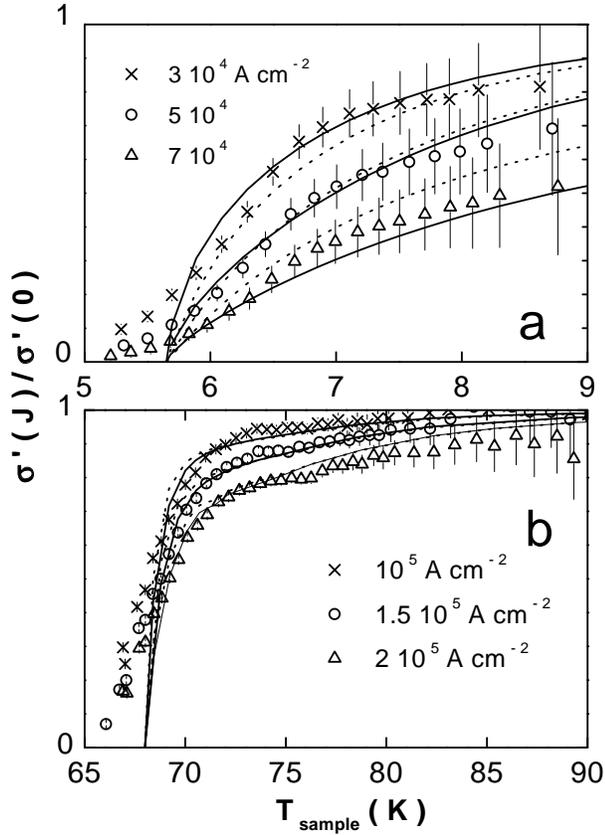}
} \caption{Sample Ia (a) and sample II (b): normalized excess
conductivity. Dotted line is Ref.~\cite{schmidt69} Gaussian
approximation using $T_c = 5.65$~K and $E_0 = 1900$ Vm$^{-1}$ for
sample Ia and $T_c = 68 $ K and $E_0 = 3\; 10^5$ Vm$^{-1}$ for
sample II. Full line is ref.~\cite{puica03} Eqs.(24),(29) , using
respectively $E_0(T_c)= 3700$~Vm$^{-1}$ and $E_0(T_c)= 4.7\;
10^5$~Vm$^{-1}$.} \label{excess}
\end{figure}

The reduced excess conductivity was obtained as :
$\sigma'(J)/\sigma'(J=0)= R(0)(R_n-R(J))/R(J)(R_n-R(0))$, where
$R_n$ is the normal state resistance. $R_n$ was obtained from the
extrapolation to below $T = 130$~K of the empirical function $R =
a+b\;T^\alpha$ for this overdoped sample\cite{kon00}, which
fitted the data in the temperature interval 130--300~K with
$\alpha=1.13$. As can be seen in Fig.~\ref{excess}, the
broadening of the transition corresponds to a drop of the reduced
excess conductivity from the normal state asymptotics
$\sigma'(J)/\sigma'(0)=1$ to zero with decreasing temperature. A
similar, but far more dramatic, behavior is observed for the
Bi$_2$Sr$_2$CuO$_{6+x}$ thin films. Samples Ia and Id were
overdoped, with $\alpha = 1.08$ and $\alpha = 1.10$ respectively.
Samples Ib and Ic were underdoped. For these samples, the normal
state resistivity was fitted using an activated law. Indeed, as
in the case of underdoped YBa$_2$Cu$_3$O$_{6+x}$\cite{tolpygo96}
and Bi$_2$Sr$_2$CaCu$_2$O$_{8+x}$\cite{kon00}, it was found that
the resistivity is well described by the activated law
$\rho_0+b\;\exp(-\Delta/T)$ below $T^*$, and by a linear
temperature dependence above. $T^*\simeq 65$~K and $T^*\simeq
85$~K was found respectively for samples Ib and Ic. As in
Ref.~\cite{tolpygo96}, it was found that $\Delta \simeq T^*$. For
all samples with $n=1$, the sample heating was found to increase
with decreasing temperature (Fig.~\ref{heating}a): this is likely
due to the decrease of the sample specific heat and of the
thermal conductivity at low temperature. For these samples, the
resistivity at temperatures well above the mid-point transition
is clearly affected by the transport current, which points to the
suppression of the excess conductivity by the current. Finally,
we stress that the resistivity curves obtained at constant $J$
are not equivalent to the constant $E$ curves usually obtained
from theoretical computations, and that the electric field
($E=R\;I\;/L$) must be calculated for each data point in order to
compare with the theoretical results.

Before we compare these results with theory, let us examine the
experimental artifacts which may affect the results given above.
First, the normal state resistivity, as extrapolated from the
high temperature values to the superconducting transition, may
not be exact. It can be shown that the error made in the reduced
excess conductivity is of the order of $(\delta R_n/\sigma')^2$.
So, the error which results from the normal state indetermination
drops strongly as one enters the superconducting transition, as
evidenced by the error bars in Fig.~\ref{excess}. Also, despite
very different normal state approximations were used (power law
or activated ones), the characteristic field obtained from the
excess conductivity was invariably well below the theoretical
one. Thus, it is unlikely that the uncertainty on the normal state
resistivity value can account for this central result of our
measurements. Then, the accuracy of the sample temperature may be
questioned. The error made in the computation of the temperature
after the main pulse is critical only for sample II. At the
middle of the transition for this sample, the broadening at the
highest current density is about 0.6~K. This is larger than the
temperature increase (0.2~K)obtained from the probe signal and,
due to the high sensitivity of the method in the transition
region, noticeably larger than the expected error on the sample
temperature. Then, the fast temperature relaxation that occurs
after the main pulse should be estimated. After correcting the
temperature, the sample resistance values above the
superconducting transition (say, 14~K and 100~K for sample Ia and
II respectively ) where the effect of current reduces to that of
heating, agree within a temperature shift of about 0.2~K. While
such a temperature difference would yield a sizeable correction
of the results for sample II, this is clearly not so for sample
with $n=1$. Finally, it may be argued that the sample
inhomogeneities affect the transition and its broadening with
electric field. The finite width of the transition ( $\simeq
8$~K, $\simeq 2$~K, $\simeq 3$~K, $\simeq 3$~K and $\simeq 5$~K
for the 10\%--90\% transition completion of samples II, Ia, Id,
Ib and Ic respectively) could be partly attributed to the
inhomogeneities that are present in the sample. While it should
be stressed that, \textit{a contrario}, a narrow transition does
not necessarily mean a more homogeneous sample and that intrinsic
fluctuations also result in a broadening of the transition (see
Fig.~\ref{resistivity}b), this could be indeed a serious
limitation in the present case. As will be seen later, the
universal behavior observed for the $n=1$ samples, independently
of their oxygen concentration, rules out such an interpretation in
this case.

The theoretical results considering only non interacting
fluctuations, i.e. Gaussian ones, for a 2D system were given in
Refs.~\cite{hurault69,schmidt69,tsuzuki69,gorkov70}. In this
case, the normalized excess conductivity is simply given by :
\begin{equation}
\sigma'(T,E)/\sigma'(T,0)=\int_0^\infty
dx\exp\{-x-[E/E_c(T)]^2x^3\} \label{schmidt}
\end{equation}
where $E_c(T) = E_0\;\epsilon^{3/2}$, $E_0 =
16\sqrt{3}k_B\;T_c/\pi\;e\;\xi_0$ and $\epsilon = (T-T_c)/T_c$.
The Gaussian result is found to account roughly for the data for
all samples (Fig.~\ref{excess}). However, the characteristic field
obtained from these fits is $E_0 \simeq 10^3$~Vm$^{-1}$ for
samples with $n=1$ and $E_0 \simeq 3\;10^5$~Vm$^{-1}$ for sample
II. These values are considerably smaller than the theoretical
estimates $E_0= 10^6$~Vm$^{-1}$ and $E_0= 3\;10^7$~Vm$^{-1}$
(using $\xi_0 = 40$~\AA{} and $\xi_0 = 20$~\AA respectively). As
shown in Ref.~\cite{puica03}, there is an enhanced suppression of
the excess conductivity, below the zero-field transition
temperature, when critical fluctuations are taken into account
within the Hartree approximation. Such an enhancement may result
in a decrease of the apparent value for $E_0$ when only Gaussian
fluctuations are considered. Within the Hartree approximation,
the magnitude of the fluctuation renormalization is essentially
determined by the Ginzburg number and the anisotropy of the
superconductivity. These parameters determine the shift of the
superconducting transition temperature with respect to the bare
mean field temperature and, associated to this shift, there is an
enhancement of the superconducting fluctuations. For sample II,
using $s= 15.35$~\AA, $\xi_{0c} = 0.2$~\AA, $\kappa = 100$
 for the superconducting plane separation, the
transverse coherence length and the Ginzburg Landau parameter
respectively, it is found ---as a result of the quasi-2D
character for this material--- that the bare mean field
characteristic temperature calculated according
Ref.~\cite{puica03,ullah91} is larger by about 30~K than the
superconducting transition temperature (in Ref.~\cite{puica03}
the fluctuating modes are cut for $k^2 > c\;\xi_0^{-2} $ and we
have used here $c=1$). Such a difference results in a large
enhancement of the superconducting fluctuations with respect to
the Gaussian approximation. This clearly fails to account for the
experimental data taken at $E=0$ (Fig.~\ref{resistivity}b).
Critical fluctuations from the Hartree approximation and the raw
superconducting parameters given above are then clearly strongly
overestimated in the region of interest for our data, and it is
not surprising that we were unable to obtain a satisfying fit of
the reduced excess conductivity in this case. However, the
Hartree approximation results were used in Ref.~\cite{livanov97}
to provide convincing fits of the resistivity data under a
magnetic field. The anisotropy parameter obtained from these fits
was $r=(2\xi_{0c}/s)^2=5\,10^{-3}$ and the Ginzburg number (using
the notation from Ref.~\cite{puica03}) was $g=G_i^2/4T_c =
4.5\,10^{-5}$~K$^{-1}$, while, from the raw superconducting
parameters given above, one obtains $r=1.7\,10^{-4}$ and
$g=6.7\,10^{-4}$ ~K$^{-1}$. These differences both contribute to
a decrease of the magnitude of the superconducting fluctuations,
and the prediction for the zero field resistivity using the same
parameter values as Ref.~\cite{livanov97} now agrees more
reasonably with the experimental data (Fig.~\ref{resistivity}b)
(although some part of the transition broadening is likely due to
the sample inhomogeneity). Following the formalism of
Ref.~\cite{puica03}, the reduced excess
 conductivity for such $r$ and $g$ values
is now essentially that of the Gaussian regime (here, $E <
5\,10^3$~Vm$^{-1}$) and the experimental data may be fitted using
$E_0(T_c)= 4.7\,10^5$~Vm$^{-1}$ (Fig.~\ref{excess}b).

\begin{figure}
\resizebox{0.5\textwidth}{!}{%
\includegraphics{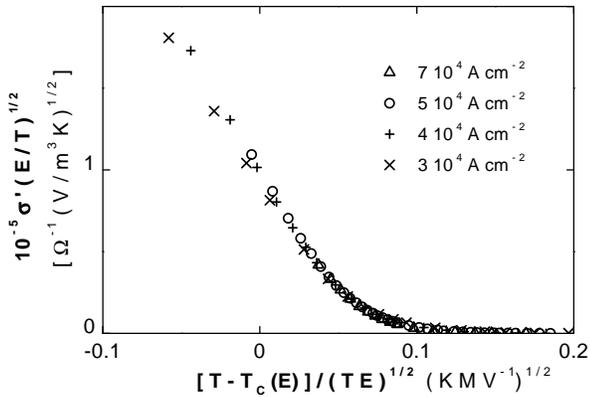}
} \caption{Sample Ia: 2D scaling, using linear $T_c(E)$.}
\label{scaling}
\end{figure}

In the case of samples with $n=1$, using the following parameters
: $s= 12.3$~\AA, $\xi_{0c} = 2$~\AA, $\kappa = 100$, the bare
mean field transition temperature is only shifted by about 0.4~K
with respect to the superconducting transition temperature. The
critical region should then be much narrower than for sample II
and, logically, the Gaussian and the Hartree approximations give
close results for $E_0$ (Fig. \ref{excess}a), i.e. well below the
above theoretical estimate for all $n=1$ samples (Table
\ref{table}). The Results of Ref.~\cite{puica03} should also be
valid also for $T < T_c$ where the normalized excess conductivity
is zero. As pointed out in Ref.~\cite{puica03}, the treatment of
the electric field dependence of the excess conductivity within
the Hartree approximation naturally yields results comparable to
those obtained in the case of a magnetic field, within the same
approximation\cite{ullah91}. As a result, one may expect that the
excess conductivity also scales with temperature and electric
field, as was observed\cite{welp91} in the case of a magnetic
field. As shown in Fig.~\ref{scaling}, the fluctuation
conductivity is found to scale according to the two-dimensional
law \cite{ullah91}: $\sigma ' \propto (T/E)^{1/2}
\mathcal{F}[(T-T_c(E))/(T\;E)^{1/2}]$, provided one takes
$dT_c(E)/dE$ a constant as given in Table \ref{table}. A
three-dimensional law ($\sigma ' \propto (T^2/E)^{1/3}
\mathcal{F}[(T-T_c(E))/(T\;E)^{2/3}]$), yields a slightly less
satisfying scaling, but comparable $dT_c(E)/dE$ value. The
critical temperature field dependence obtained in this way
confirms the fit of the reduced excess conductivity with the
results in ref.\cite{puica03}, as the critical field $T_{c0}
(dT_c/dE)^{-1}$ is also well below the theoretical estimate (Table
~\ref{table}).

Thus, there are converging experimental pieces of evidence that
the apparent characteristic field ---at least for samples with
$n=1$, for which the transition broadening is much larger than the
error possibly made for the sample temperature--- is well below
the one expected from a simple estimate of the depairing field.
Several explanations may be considered for this. First, it is
known that the existence of 'hot spots', where the sample
temperature locally exceeds $T_c$, may drive the sample into the
normal state. Such instabilities must be ruled out above $T_c$,
i.e. in the analysis of the reduced conductivity. Below $T_c$, the
characteristic current density for which hot spots develop may be
estimated to be \cite{skocpol74}
$10^{10}\;(T_c-T)^{1/2}$~Am$^{-2}$, well above current densities
which significantly affect the resistive curve of $n=1$ samples.
Then, as noticed above, the transition width of the sample is
finite, which may be interpreted as the result of a distribution
of the superconducting temperature. The effect of $T_c$
inhomogeneities on the superconducting transition is complicated
and the full percolation problem (with a non-linear conductivity)
should be considered in the general case. The situation is more
simple when the resistivity is close to the normal state value,
as the assumption of a uniform current density may be used. Using
this assumption in the case of the larger current displayed in
Fig.~\ref{resistivity}a and the results of Ref.~\cite{puica03},
it is found that the resistivity curve for a field $E/E_0\simeq
10^{-4}$, as would be expected from the theoretical estimates
above, is only weakly affected by a distribution of $T_c$ with a
typical width of 1.5~K. In addition, within the hypothesis that
this $T_c$ distribution strongly influences the apparent value
for the characteristic electric field, one would expect that this
parameter would be highly sensitive to the oxygen doping state.
This is clearly not so, as demonstrated in the case of the $n=1$
samples (Table~\ref{table}): although the optimal doping might
coincide with a maximum in $E_0$, the characteristic electric
field is well below the depairing estimate for all samples.
Another possible explanation for the apparent weakness of the
critical field is that the observed current-induced resistivity
is the one of the intergranular material. Such a contribution is
unlikely for these epitaxial films where grains (typically 0.1
$\mu$m large) are well oriented by the substrate and exhibit sharp
(0$^\circ$--90$^\circ$) grain boundaries at the atomic level. It
is known also that granular bulk or thin films materials close to
$T_c$ are invariably in the Ginzburg-Landau regime, where the
critical current is the one needed for the suppression of the
order parameter in the grains\cite{darhmaoui96}. Moreover, this
would require that the intergranular weak links contribution to
the resistivity is comparable to the normal state value, which
would likely imply a two step transition curve (corresponding to
the Ginzburg-Landau to the Ambegaokar-Baratoff behavior crossover
temperature) at large currents, which is not observed here. The
scaling of the data according to the prediction of the depairing
mechanism rather suggests that a microscopic mechanism (at the
coherence length scale) should be found. Microdomains at this
scale or below\cite{pan01} is one of such candidates and it has
already been noticed that they may affect the magnitude of the
depairing current\cite{darhmaoui96}. Along this line, it is worth
mentioning that the $n=1$ material also exhibits a peculiar
behavior with respect to the superconducting Nernst effect.
Indeed, whereas the Nernst effect for YBa$_2$Cu$_3$O$_{6+x}$ and
Bi$_2$Sr$_2$CaCu$_2$O$_{8+x}$ shows a peak which shifts with
magnetic field to lower temperature in a way similar to the
resistive transition\cite{ri94}, the underdoped $n=1$ material has
shown a nearly field independent peak\cite{capan03}. Moreover, the
Nernst effect clearly has its maximum well above the resistive
transition temperature (i.e. in the fluctuation regime), whereas
this peak points ---at zero field--- towards the completed
resistive transition in the case of the two former compounds. As
underlined in Ref.~\cite{capan03}, this discrepancy between the
resistive and the Nernst effect transitions may be due to an
anomalously large vortex contribution, such as the one arising
from a vanishing vortex viscosity. Such a mechanism could account
for an enhanced dissipation with current, as observed here. The
existence of a phase coherence length distinct from the amplitude
one, as proposed in the case of microscopic
inhomogeneity\cite{pan01}, may also account for these anomalies.
Finally, the contribution of the $d$-wave symmetry to the
depairing field should also be considered as, in the nodal
directions, the depairing field is virtually zero: although there
is to our knowledge no theoretical evaluation in this case, we
expect a decrease of the critical field with respect to the
conventional $s$-wave symmetry, as well as an in-plane angular
dependence of the non-linear resistivity. Clearly, further
studies of the effect of large electric field, in particular on
compounds which have shown distinct resistive and thermodynamic
lines ---such as
Tl$_2$Ba$_2$CuO$_{6+\delta}$\cite{carrington96}--- are needed to
clarify these mechanisms.

\begin{acknowledgments}
We acknowledge the support of CMCU to project 01/F1303.
\end{acknowledgments}
% BibTeX users please use
% \bibliographystyle{}
% \bibliography{}

\newpage

\begin{table}
\caption{\label{table}Sample parameters}
\begin{ruledtabular}
\begin{tabular}{cccccc}
 Sample&n&$T_c (K)$&doping&$E_0$\footnotemark[1]&$T_c(dT_c/dE)^{-1}$\footnotemark[2]\\
&&(K)&&$V m^{-1}$&$V m^{-1}$\\
\hline II& 2 & 67.4 & overdoped &$4.7\;10^5$&- \\
% 130201mmd
Ia& 1 & 5.65 & overdoped &$3.7\;10^3$&$9.5\;10^2$\\
% 130201mmg
Ib& 1 & 9.8 & underdoped\footnotemark[3] &$7.8\;10^3$&$1.5\;10^3$\\
% 140201mm
Ic& 1 & 14.4 & underdoped &$1.5\;10^4$&$4.8\;10^3$\\
% 140201mb
Id& 1 & 5.1 & overdoped &$1.8\;10^3$&$3.8\;10^2$\\
\end{tabular}
\end{ruledtabular}
\footnotetext[1]{from the fit using the results of
\cite{puica03}.} \footnotetext[2]{from the 2D scaling.}
\footnotetext[3]{same sample as Ia after vacuum annealing}

\end{table}

\end{document}